\documentclass[review]{elsarticle}

\usepackage{graphicx}

\graphicspath{{./pdf/}{./eps/}}

\DeclareGraphicsExtensions{.pdf,.jpeg,.png,.eps}

\usepackage{ amssymb }
 \usepackage{subfigure}%
\usepackage{hyperref}
\usepackage{algorithmic}
\usepackage[linesnumbered,ruled,vlined]{algorithm2e}	
\SetAlFnt{\footnotesize}
\SetAlCapFnt{\footnotesize}
\SetAlCapNameFnt{\footnotesize}
\SetKwFor{While}{when}{:}{fintq}%

\journal{}

\begin{document}

\begin{frontmatter}

\title{SITAN: Services for Fault-Tolerant Ad Hoc Networks with Unknown Participants}

\author[mymainaddress]{David~R.~Matos}
\author[mysecondaryaddress]{Nuno~Neves}
\author[mysecondaryaddress]{Alysson~Bessani}



\address[mymainaddress]{INESC-ID, Instituto Superior T\'{e}cnico, Universidade de Lisboa, Portugal}
\address[mysecondaryaddress]{LaSIGE, Faculdade de Ci\^{e}ncias, Universidade de Lisboa, Portugal}

\begin{abstract}
The evolution of mobile devices with various capabilities (e.g., smartphones and tablets), together with their ability to collaborate in impromptu ad hoc networks, opens new opportunities for the design of innovative distributed applications. The development of these applications needs to address several difficulties, such as the unreliability of the network, the imprecise set of participants, or the presence of malicious nodes. In this paper we describe a middleware, called SITAN, that offers a number of communication, group membership and coordination services specially conceived for these settings. These services are implemented by a stack of Byzantine fault-tolerant protocols, enabling applications that are built on top of them to operate correctly despite the uncertainty of the environment. The protocol stack was implemented in Android and NS-3, which allowed the experimentation in representative scenarios. Overall, the results show that the protocols are able to finish their execution within a small time window, which is acceptable for various kinds of applications. 
\end{abstract}


\end{frontmatter}


\section{Introduction}

In the last years there has been a proliferation of mobile devices: laptops are increasingly lighter and powerful, smartphones and tablets are becoming much more common and a new type of hybrid computers, combining the mobility of tablets and the processing capabilities of laptops are now a reality. All these devices share a group of characteristics that sets them apart: wireless communication, a good level of mobility, and some constrains on processing, memory and battery power. 

There is a range of distributed applications that can be developed for this kind of devices, such as: productivity tools for groups, schedulers, chat, save and rescue support for emergency scenarios, and multimedia sharing~\cite{ex1,ex2,ex3}. These distributed applications take advantage of the wireless communication capacity and mobility of the devices, but must deal with the limitations of the environment: unreliability of the communication medium, due to noisy channels or obstacles to messages transmission; unreliability of the devices, as they may disconnect at any time or be apart from the network; asynchrony of message delivery; and the potential for different forms of attacks.
 


This paper presents SITAN, a middleware that offers services for the implementation of applications in  mobile ad hoc networks (MANET). These networks are self organized because there is no centralized unit to configure and manage them. Three major challenges had to be addressed in the design of SITAN. \emph{Lack of knowledge} is something inherent of this type environment, and it is perceived by having an inaccurate view of which nodes\footnote{Throughout the paper we will use interchangeably the words processes, nodes and devices.} are present in the network and how they are connected to each other. Moreover, since nodes are mobile, the knowledge about the network organization can become outdated relatively fast. Another problem of MANETs is the fact that wireless communications, due to their exposure and simplicity of access, are vulnerable to many \emph{distinct forms of attack}. Therefore, it is advisable to consider that potentially some nodes may become compromised by a malicious adversary (i.e., they suffer a Byzantine failure) and that parts of the network may also misbehave. Lastly, in order for the middleware to be useful, it should be \emph{efficient}. This means that protocols should take advantage of the characteristics of MANETs, such as the ability to broadcast messages to a node's neighbors with little cost.

SITAN is organized as a stack of protocols, which are divided in three main layers: communication, group membership and agreement (see Figure~\ref{fig:stack}). The protocols were in some cases adapted and optimized from previous proposals, while others are novel. All protocols were implemented from scratch. The two lower layers follow the organization described by Alchieri et al.~\cite{bft-cup}. A straightforward implementation of the layers had however several limitations in performance, and consequently it was necessary to redesign some of the protocols to leverage from the broadcast capability of the wireless medium. 

The lack of a central coordinator to configure and organize the network hampers its management. The distributed applications and some of the management operations require several coordination actions, such as: selecting a unique process to be leader; deciding the ordering of the received messages; assigning roles to processes; deciding routes. To facilitate the implementation of these operations, SITAN offers three kinds of agreement primitives. 
Binary consensus is based on the Turquois~\cite{turquois}. This protocol was proposed for resource-constrained devices and it circumvents two impossibility results relevant to MANETS~\cite{flp,sw}. The multivalued and vector consensus~\cite{Correia11} protocols are novel, and they support agreement on more complex values.

SITAN was implemented both in devices running Android and in the NS-3 simulator~\cite{NS3}, which were used to run several experiments. Results show that the protocols have a good level of performance, with acceptable latencies for various kinds of applications. Less favorable environmental conditions have a limited impact on performance, but we expect that this is a small penalty for the ability to communicate with partial connectivity and tolerate different failures (in the nodes and the network). 

This work has the following main contributions: (1) it proposes SITAN, a novel protocol stack with communication and coordination services particularly tailored for MANET; (2) it presents two new protocols for multivalued and vector consensus, which exhibit a high scalability accordingly to simulations; (3) the  evaluation of SITAN with Android devices and the NS-3.

%
%
%
%
\section{Related Work}\label{sec:rel_work}

The consensus problem abstracts the capability to reach agreement among a group of processes operating in a distributed system~\cite{turek1992many}. Solving consensus allows processes to coordinate their actions and maintain a coherent state despite the difficulties caused by the environment, such as failures and unpredictable delays.  In such settings, however, consensus is hard to achieve because it is bound by several impossibility results. For example, FLP~\cite{flp} states that it is impossible to solve consensus in a deterministic way in an asynchronous system if only one process (crash) fails. Another relevant impossibility result, which is due to Santoro-Widmayer~\cite{sw}, is related with systems in which failures may occur in the communication links.  Therefore, over the years, consensus has been extensively studied in conventional networks, where solutions have explored different mechanisms (e.g., randomization) to circumvent the existing limitations (as described by the following survey papers~\cite{turek1992many,Correia11,Tseng2016}).

In MANETs solving the consensus problem is particularly complex. Several aspects contribute to this, including the unpredictability of message deliveries, the ease for a malicious actor to access the communication medium, and the dynamic set of processes that might be present at each instant. With respect to this last point, the \textit{consensus with unknown participants} (CUP)~\cite{cup} has been studied, in order to find the conditions necessary for solving consensus in a network in which the (failure-free) participants are not known \textit{a priori}.  In an extended work~\cite{DavidCavin2005} the authors go further and solve the fault-tolerant CUP (FT-CUP), which assumes a network where some participants may crash or leave. An improved solution for FT-CUP is presented in~\cite{greve2007knowledge}, where they analyze the relation between knowledge connectivity and synchrony for consensus. More recently, there has been a proposal for solving consensus with Byzantine failures in asynchronous systems~\cite{greve2010,bft-cup}.

Previous work has mostly focused on finding the necessary and sufficient conditions for solving the CUP problem. To our knowledge, SITAN is the first practical work that supports MANET applications by leveraging from recent results in the area of CUP. It also includes two new consensus protocols that enable agreements on more complex values. An implementation was developed, allowing the experimentation of these protocols in realistic scenarios, under favorable and adverse conditions.

\section{System model}\label{sec:sysmodel}

The system is composed by a group $\Pi$ of processes (also called participants or nodes) belonging to larger universe $U$. Since the network is created in an ad hoc manner, a process $i \in \Pi$ may only be aware of a subgroup $\Pi_i \subseteq \Pi$. A process $i$ can only send a message directly to another process $j$ if $j \in \Pi_i$, i.e., if $i$ knows $j$. As expected, if $i$ receives a message from $j$ and $j \not \in \Pi_i$, then $i$ will add $j$ to $\Pi_i$ (i.e., $i$ becomes acquainted to $j$ and can send messages to it). 

All processes have a unique identifier and a pair of public ($pu_i$) and private ($pr_i$) keys. Processes attempt to cache a copy of the public key of the other processes, but in case it is not available, the owner of the public key can always append it to its messages. Every process has the ability to check the validity of public keys of other processes (e.g., public keys are placed in a certificate signed by a well known entity). Public/private keys are used to establish session keys in a secure way, which then support the creation of authenticated reliable (point-to-point or multicast) channels. In addition, they are used to sign values, allowing their validation by other processes.

Processes can suffer Byzantine faults, which means that they can become silent, send messages with wrong values, or work together with other faulty processes to corrupt the system. It is assumed a bound $f$ on the number of faulty processes in $\Pi$. The protocols operate correctly as long as $n \geq 3f+1$ (where $n$ is the total number of nodes in $\Pi$). The fault model also assumes that the network can experience Byzantine faults. For example, it can have omissions, reorder or corrupt messages.

With regard to synchrony, the system is assumed to be asynchronous. All protocols make progress at the speed that the network delivers messages, meaning that if the network is slow they will take longer to finish (e.g., if the network is under attack, processes will continue to retransmit messages until their eventual arrival, allowing the protocols to complete). Under these conditions, the binary consensus protocol is able to terminate with very high probability by employing a randomization technique~\cite{turquois}.

%
%
%
%
\section{Protocol stack}\label{sec:stack}

\begin{figure}
\centering
\includegraphics[scale=0.45]{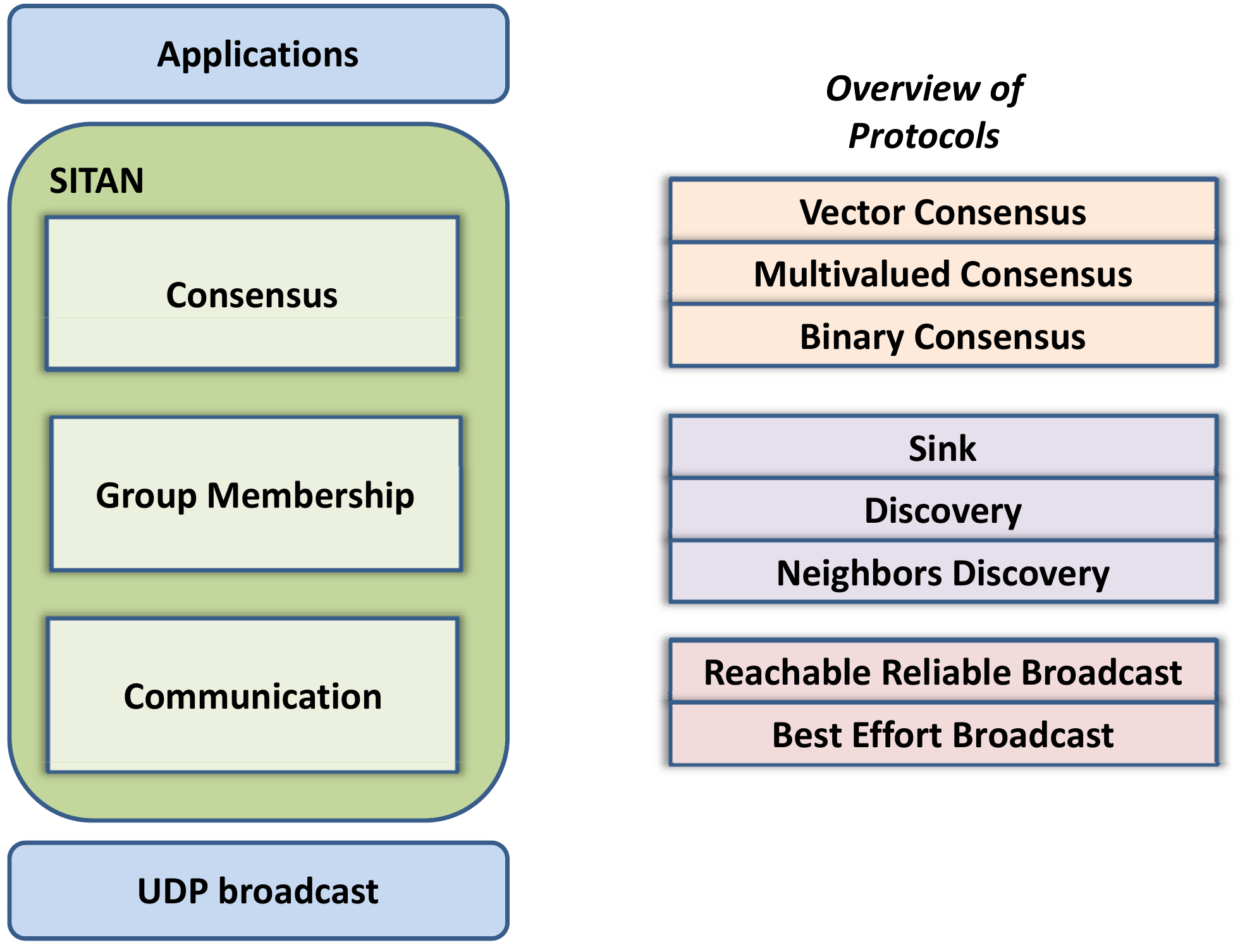}
\caption{SITAN protocol stack}
\label{fig:stack}
\end{figure}

The SITAN protocol stack is divided in three main layers (see Figure~\ref{fig:stack}), providing services that facilitate the development of distributed applications by hiding most of the complexity inherent of MANET environments. The layers offer fundamental services related to communications, group membership and agreement, and they were organized to achieve two main objectives: (i) 
applications can call operations in any of three layers to take advantage of the available services; (ii) services above can  use any operations provided below. In the rest of this section, we will describe the protocols that implement the various services.

\subsection{Communications layer}\label{sec:comm_layer}
The communication layer employs the underlying broadcast primitive (UDP) of the mobile devices to disseminate values, allowing several processes to be reached with a single message transmission. Depending on the configuration and the device capabilities, it can be implemented with a WIFI or WIFI Direct channel. The Wifi channel requires an available access point.

This layer includes two broadcast services: a simpler and less reliable \textit{best effort broadcast} and a more complex \textit{reachable reliable broadcast}. 

\paragraph*{Best Effort Broadcast}\label{sec:beb}
This protocol simply transmits messages securely to the immediate neighbors of a source $p_i$. Since it is based on UDP, no assurances are given about successful delivery, as no mechanisms are employed to re-transmit or forward messages. It however takes advantage of signatures and sequence numbers to ensure that messages are only delivered once (no duplicates) and that only recognized senders have their messages accepted (no impersonation). This service is used, for instance, by the \textit{neighbors discovery} and \textit{discovery} protocols to identify the neighbors of $p_i$ and to start the construction of a network map. 



\paragraph*{Reachable Reliable Broadcast}\label{sec:rb}
Allows a process $p_i$ to broadcast a message reliably to every node that can be $reached$ in the ad hoc network. Every message is forwarded by every node until it arrives to the destinations. Retransmissions are performed until reception is acknowledged.  Each message contains data about the nodes that were visited. As side effect, when the message arrives at a process, it can assist on the construction of a graph with the locally perceived knowledge (i.e., which nodes are known to other nodes). Notice however that this graph is potentially incomplete and distinct among nodes (e.g., some connections may be missing because they go through a Byzantine node that drops packets). This protocol uses the best effort broadcast, guaranteeing that the message is only delivered if it was received by at least $f + 1$ independent paths. 


\subsection{Group membership layer}\label{sec:gm}

The group membership layer is responsible for keeping track of the active processes in the network and compute the network graph. 
It contains three protocols: both \textit{neighbors discovery} and \textit{discovery} were implemented following the description  in~\cite{bft-cup}, however, \textit{sink} had to be modified to match our environment.

\paragraph*{Neighbors discovery}\label{sec:pd}
This is the first protocol to run when the group membership layer starts, and its main purpose is to find the neighbors of $p_i$. It has an internal state containing a list of tuples with the identifiers of the neighbors and the last time it heard from them,  $ neighbors_i = \{ \langle j, last\_heart\_beat_j \rangle \}$. A single instance of this protocol runs in background during the entire execution of SITAN. It  works by periodically transmitting heartbeats through best effort broadcast. When a new neighbor $j$ is detected then the set is updated to $neighbors_i \gets neighbors_i \cup \langle j, currentTimestamp \rangle$. A timeout is triggered periodically to check if all processes sent a heartbeat. A process with missing heartbeats is excluded from $neighbors_i$ and the sink protocol is notified to check if this process was in the sink. If so, a new sink protocol execution is scheduled.

\paragraph*{Discovery}\label{sec:discovery}
This protocol aims to draw the graph of the network. For a process $p_i$, it contains: an internal state with the set of known processes, initialized with the $neighbors_i$  provided by the neighbors discovery protocol; a list of tuples with the neighbors of each known node ($nnei_i$) and a list of processes that did not send their neighbors list to $p_i$ ($pending_i$). When the protocol starts, processes ask their neighbors for their neighbors list using the \textit{GET\_NEIGHBORS} message. When a process receives this request, it answers with a \textit{SET\_NEIGHBORS} response including its neighbors list. Finally, when process $p_i$ gets the response from $p_j$, $p_i$ updates its $nnei_i$ to include the information from $neighbors_j$ and removes $j$ from $pending_i$. When the protocol finishes, each process has a list with the known processes ($known_i$).

\paragraph*{Sink}\label{sec:sink}
The protocol finds a subgroup of processes  that has sufficient connectivity to run consensus, which is called the \textit{sink}. This group fulfills a set of requirements, namely: ensures a minimum size of the group to tolerate a certain number of failures; all processes in the group are known to each other; and every process is reachable without requiring message forwarding.
%





This protocol is a modified version of the one presented in~\cite{bft-cup} in order to reduce the number of exchanged messages and the number of steps for completion. In the original design, $p_i$ would reachable broadcast its set of known nodes ($known_i$). When $p_j$ receives $known_i$ it would respond to $p_i$ confirming if the $known_i$ and $known_j$ were similar. If so they would include each other in the sink group. This implementation used messages that were sent  in a point to point manner which is not efficient because it does not leverage from the broadcast facility provided by the wireless medium. Also, it would require a two step process to compare known nodes. The new version of protocol leverages from the reachable reliable broadcast to minimize the communications as several processes are contacted in every transmission.


\subsection{Consensus layer}\label{sec:consensus_layer}
Consensus protocols allow processes to propose a value to the group and then decide on a common result. The layer solves three versions of consensus: \textit{binary} lets processes agree on a single bit value (e.g., true or false); \textit{multivalued} enables agreements on arbitrary values; and \textit{vector} returns an agreed upon vector with proposal values from a subset of the processes.

In our implementation, the consensus protocols are run only by the processes that are included in the \textit{sink}. When a consensus execution ends, the processes in the \textit{sink} broadcast the result to the remaining nodes in the network. This improves efficiency and scalability because less processes are involved, contributing for a decrease on the network traffic.

\paragraph*{Binary Consensus}\label{sec:bc}
The service is implemented by a randomized $k$-consensus protocol. Every process $p_i$ proposes a value $v_i$, such that $v_i \in \{0,1\}$, and decides on a value $v \in \{0,1\}$. Only $k$ processes, such that $\frac{n+f}{2} < k \leq n - f $, are expected to decide. The remaining correct processes (at most $n-k$) might not decide, but if they do, the same value must be selected as the rest. The protocol ensures the following properties:

\begin{itemize}
\item \textbf{BC1 Validity.} If all processes propose the same value $v$, then any correct process that decides, decides $v$.
\item \textbf{BC2 Agreement.} No two correct processes decide differently;
\item \textbf{BC3 Termination.} At least $k$ correct processes eventually decide with probability 1.
\end{itemize}

The randomized protocol we use is fully described in~\cite{turquois}, and therefore in the interest of saving space we only give a short overview. The protocol makes progress in rounds of three phases: in the \textit{CONVERGE} phase, processes pick the most common proposal value; in the \textit{LOCK} phase, each process chooses a value that was selected by a sufficiently large quorum of nodes (greater than $\frac{n+f}{2}$) or $\bot$ if no such quorum was found; in the \textit{DECIDE} phase, processes confirm if there is a quorum with the same value in order to finish consensus, otherwise a new round begins (possibly with a random value as proposal).





\paragraph*{Multivalued Consensus}\label{sec:mvc}
This service allows processes to agree on a value of arbitrary length. Our algorithm solves the multivalued consensus problem in the presence of Byzantine faults in an ad hoc network. The following properties are ensured:
\begin{itemize}
\item \textbf{MVC1 Validity.} If all correct processes propose the same value $v$, then any correct process that decides, decides $v$.
\item \textbf{MVC2 Validity.} If a correct process decides $v$, then $v$ was proposed by some process or $v = \bot$.
\item \textbf{MVC3 Validity.} If a value $v$ is proposed only by corrupt processes, then no correct process that decides, decides $v$.
\item \textbf{MVC4 Agreement.} No two correct processes decide differently.
\item \textbf{MVC5 Termination.} Every correct process eventually decides.
\end{itemize}

\begin{algorithm}
\KwIn{Unique instance identifier of this execution $mid$}
\KwIn{Initial proposal value $proposal_i$}
\KwOut{Decision value $v_i$}
$\phi \gets 0$\; \label{mvc:1}
$v_i \gets proposal_i$\;\label{mvc:2}
$M_i \gets 0$ \;\label{mvc:2b}

\textbf{Task T1} \\\label{mvc:3}
\While{local clock tick}{\label{mvc:4}
	$J_i \gets$ justification for $v_i$ based on messages in $M_i$\;\label{mvc:5a}
    $reachableBroadcast(\langle mid, i, \phi, v_i, J_i, sig_i \rangle)$\;\label{mvc:5}
}

\textbf{Task T2} \\\label{mvc:6}
\While{$m = \langle mid, j, \phi_j, v_j, J_j, sig_j \rangle$ is received}{ \label{mvc:7}
	$M_i \gets M_i \cup \{m:m$ $is$ $valid\}$\;\label{mvc:8}
}

\textbf{Task T3} \\\label{mvc:9}
\While{$|\{ \langle mid, *, \phi, *, *, * \rangle \in M_i: \phi = 0\}| > \frac{n+f}{2}$}{ \label{mvc:10}
	$maj \gets majority$ $value$ $in$ $M_i$ $for$ $\phi = 0$\;\label{mvc:11}
	\If{$|\{ \langle mid, *, \phi, v, *, * \rangle \in M_i: \phi = 0 \wedge v = maj \}| > f$}{\label{mvc:12}
		$v_i \gets maj$\;\label{mvc:13}
	}
	$\phi \gets 1$\;\label{mvc:14}
}
\While{$|\{ \langle mid, *, \phi, *, *, * \rangle \in M_i: \phi = 1 \}| > \frac{n+f}{2}$}{\label{mvc:15}
	\eIf{$|\{ \langle mid, *, \phi, v, *, * \rangle \in M_i: \phi = 1\}| > \frac{n+f}{2}$}{\label{mvc:16}
		$v_i \gets v$\;\label{mvc:17}
		$propose \gets 1$\;		\label{mvc:18}
	}{\label{mvc:19}
		$v_i \gets \bot$\;\label{mvc:20}
		$propose \gets 0$\;\label{mvc:21}
	}
	$result \gets BConsensus(propose)$\;\label{mvc:22}
	\eIf{$result = 1$}{\label{mvc:23}
		\If{$v_i = \bot$}{\label{mvc:24}
			\textbf{wait until} (it is delivered $\langle mid, *, \phi, v, *, * \rangle \in M_i:\phi = 2$)\\{\label{mvc:25}
				$v_i \gets v$\;\label{mvc:26}
			}
		}
	}(\tcc*[f]{$result = 0$}){  \label{mvc:27}
		$v_i \gets \bot$\;\label{mvc:28}
	}
	$\phi_i \gets 2$\;\label{mvc:29}
}
\Return $v_i$\;\label{mvc:30}
\caption{Multivalued consensus protocol}
\label{alg:mvc}
\end{algorithm}

Algorithm \ref{alg:mvc} solves the multivalued consensus problem. Each process $p_i$ starts with its own value $proposal_i$ and the same identifier for this consensus execution $mid$. Then, it maintains throughout the execution an internal state with: the current phase number $\phi$, the current value for the result $v_i$ and a set with the received messages $M_i$. The algorithm works with three independent tasks.

Task T1 is periodically triggered by a local timer (line \ref{mvc:4}). It reachable broadcasts the internal state of $p_i$ in a message with $\langle mid, i, \phi, v_i, J_i, sig_i \rangle$, where $i$ is the identifier of the sender and $\phi$ is the current phase number. $J_i$ is a set of messages that justifies the proposal of value $v_i$ of the process $p_i$ (line \ref{mvc:5a}). In phase $\phi = 0$, for instance, this set is empty as $p_i$ simply transmits its own proposal (other examples are given below). The message also includes a signature $sig_i$ of its content for protection against tampering.

Task T2 handles the arriving messages. When a message is received, it is validated and, if correct, it is stored in $M_i$ (lines \ref{mvc:7} -- \ref{mvc:8}). The validation includes two main tests: (i) the signature is checked to prevent impersonation and malicious changes in the contents; and (ii), the verification that the included value is justified by the  $J_j$ set of messages. This last test ends up constraining Byzantine processes to transmit values that are acceptable accordingly to the protocol execution (otherwise, their messages will be simply dropped).

Task T3 updates the internal state depending on the phase $\phi$:
\begin{itemize}
\item $\phi = 0$: once the process collects a quorum of more than $\frac{n+f}{2}$ messages with this phase, it sets a variable $maj$ with the most proposed value (lines \ref{mvc:10} -- \ref{mvc:11}). Then, if at least one correct process voted for $maj$ (line \ref{mvc:12}), it is safe to update the local value $v_i = maj$.  By performing this update, the local process decides to go with the majority, thus increasing the probability that in the end the result of consensus is different from $\bot$. Finally, the phase number $\phi$  is incremented (line~\ref{mvc:14}).

\item $\phi = 1$: the process waits until a quorum of $\frac{n+f}{2}$ messages for this phase has arrived (line \ref{mvc:15}). Then, if there is a significant number of messages with the same value, the local value $v_i$ can be updated (lines \ref{mvc:16} -- \ref{mvc:18}).  Otherwise, $v_i$ is set to $\bot$  (lines \ref{mvc:19} -- \ref{mvc:21}). Notice that quorum $\frac{n+f}{2}$ ensures that correct processes either pick the same value $v$ or they set their local value to $\bot$. The $propose$ variable is updated accordingly, and is used as proposal value for the binary consensus (line \ref{mvc:22}). The decision of the binary consensus is then used to determine the outcome of the protocol:
\begin{itemize}
\item $result = 1$: there was agreement in a proposal by correct processes. In this case, however, it can happen that $p_i$ was unable to select that value before the binary consensus (i.e., $v_i = \bot$). Therefore, it must wait until one of the other processes sends the decided value (lines \ref{mvc:25} -- \ref{mvc:26}). Notice that a malicious process $p_j$ would not be able to trick $p_i$ into accepting a wrong value. Recall that the message containing this value would need to include a justification $J_j$ with more than $\frac{n+f}{2}$ messages of $\phi = 1$ with that erroneous value, which would be impossible to forge because correct processes follow the algorithm;
\item $result = 0$: processes could not agree in a proposal by a correct process. In this case, $\bot$ will be the chosen value (lines \ref{mvc:27} -- \ref{mvc:28}).
\end{itemize}
To finish, phase number $\phi$ is incremented and the decision value $v_i$ is returned (lines \ref{mvc:29} -- \ref{mvc:30});
\item $\phi = 2$: in background, Task T1 continues to broadcast the decided value to ensure that the other processes can also terminate (if they are blocked in line~\ref{mvc:25}).
\end{itemize}

\paragraph*{Vector Consensus}\label{sec:vc}
In the vector consensus problem each process $p_i$ proposes a value and all correct processes decide on a common vector $V$ of size $n$. $V[i]$ contains the value proposed by $p_i$ or $\bot$. The following properties are ensured:
\begin{itemize}
\item \textbf{VC1 Vector Validity.} Every correct process that decides, decides on a
vector $V$ of size $n$:
\begin{itemize}
\item $\forall_{p_i}$, if $p_i$ is correct, then either $V[i]$ is the value proposed by $p_i$ or $\bot$;
\item At least $(f+1)$ elements of $V$ were proposed by correct processes.
\end{itemize}
\item \textbf{VC2 Agreement.} No two correct processes decide differently.
\item \textbf{VC3 Termination.} Every correct process eventually decides.
\end{itemize}

\begin{algorithm}
\KwIn{Unique instance identifier of this execution $vid$}
\KwIn{Initial proposal value $proposal_i$}
\KwOut{Decision vector $V_i$}
$r \gets 0$\; \label{vc:1}
$array_i \gets ((\bot, ..., \bot), ..., (\bot, ..., \bot))$\;\label{vc:2}
$array_i[i][i] \gets \langle proposal_i, sig_i \rangle$\;\label{vc:3}
\textbf{Task T1} \\\label{vc:4}
\While{local clock tick}{\label{vc:5}
	$reachableBroadcast(\langle vid, i, array_i[i] \rangle)$\;\label{vc:6}
}

\textbf{Task T2} \\\label{vc:7}
\While{$m = \langle vid, j, vector_j \rangle$ is received}{\label{vc:8}
	$count \gets 0$\; \label{vc:9}
	\ForAll{$(v_k, sig_k) \in vector_j$}{		\label{vc:10}
		\If{$v_k \neq \bot$}{\label{vc:13}
			\If{$verifySig(k, v_k, sig_k) = invalid$}{\label{vc:11}
				\textbf{continue to when}\; \label{vc:12}
			}
			$count \gets count + 1$\;\label{vc:14}
		}		
	}
	\If{$count = 2f+1$}{\label{vc:15}
		$array_i[j] \gets vector_j$\;\label{vc:16}
	}
	\If{$| \{ v_k = array_i[i][k]: k \in \{0, ..., n-1 \} \wedge v_k \neq \bot \}| < 2f+1$}{\label{vc:17}
		$array_i[i][j] \gets vector_j[j]$\;\label{vc:18}
	}
}
\textbf{Task T3} \\\label{vc:19}
\While{$\exists array_i[j]: |\{v_k = array_i[j][k] : k \in \{0,...,n-1\} \wedge v_k \neq \bot\}| = 2f + 1 $}{ \label{vc:19a}
	\Repeat{$V_i \neq \bot$}{\label{vc:20}
		\For{$index \in \{0, ..,n-1\}$}{\label{sv:2}
			$j \gets (r + index)$ mod $n $\; \label{sv:2a}
			\If{$|\{v_k = array_i[j][k] : k \in \{0, ..., n-1\} \wedge v_k \neq \bot \}| = 2f+1$}{ \label{sv:3}
        \textbf{break}\; \label{sv:3a}
	}	
}

        $mid \gets \langle vid, r \rangle$\;\label{sv:22a}
		$V_i \gets MConsensus(mid, array_i[j])$\;\label{vc:22}
		$r \gets r+1$\;\label{vc:23}
	}\label{vc:24}
}
\Return $V_i$\;\label{vc:25}
\caption{Vector consensus protocol}
\label{alg:vc}
\end{algorithm}

Algorithm \ref{alg:vc} solves the vector consensus problem. Every process $p_i$ starts with the same identifier for the current consensus instance $vid$ and a proposal value $proposal_i$. The internal state that is maintained is composed by a $n \times n$  array $array_i$ and the current round number $r$. Each row $j$ of the array $array_i[j]$ corresponds to a vector constructed by process $p_j$. The entry $i$ of the local vector of process $p_i$ is initialized with $proposal_i$ together with its signature (line \ref{vc:3}). The protocol works with three tasks.

Task T1 is triggered by a local clock to broadcast a message with the vector  that is being built locally $array_i[i]$ (lines \ref{vc:5} -- \ref{vc:6}). 

Task T2 handles the received messages. When $p_i$ gets a message sent by $p_j$ (line \ref{vc:8}), it checks that the included vector $vector_j$ is valid (lines \ref{vc:9} --\ref{vc:14}). To accomplish this, it validates the signature associated with each value in the vector (with function $\mathit{verifySig(k, v_k, sig_k)}$ that checks the digital signature of $v_k$ by using the public key $pu_k$ of process $p_k$). If one of the signatures is invalid, the processing is interrupted and the message is discarded (line \ref{vc:12}). In the end, if the number of included values is equal to $2f+1$ then the vector is  stored locally in the array because it fulfills the requirements for a valid result of  consensus (property VC1 and lines \ref{vc:15} -- \ref{vc:16}). If the number of entries in the local vector $array_i[i]$ is less than $2f + 1$, then it is updated to include the proposal of $p_j$ (lines \ref{vc:17} -- \ref{vc:18}).

Task T3 selects the final result. It is triggered when there is a vector with $2f+1$ values different from $\bot$ (line~\ref{vc:19a}). Then, it enters in a loop where it attempts to agree in one of the proposed vectors using the multivalued consensus. In each iteration, it selects one of the vectors stored in the array, ensuring that it contains the necessary $2f+1$ entries (lines~\ref{sv:2} -- \ref{sv:3a}). Notice that at least one  vector always exists because the $when$ clause was activated (line~\ref{vc:19a}). In addition, since the search for the vector starts in a different row of the array, as it depends on the round number $r$ (in line~\ref{sv:2a}, variable $j = r\ mod\ n$ when $index=0$), all included vectors will eventually have the opportunity of being selected. Then, the multivalued consensus is called to decide on a vector (line~\ref{vc:22}). When the result of the multivalued consensus is different from $\bot$, vector consensus can terminate with the returned value (line~\ref{vc:25}).


%
%
%
%
\section{Implementation}
SITAN was implemented as a library that can be integrated in MANET applications. The library includes the various services for $p2p$ communication, group membership and consensus that were explained. It can be configured to operate with the WiFi protocol, requiring an access point, or with WiFi Direct in a pure $p2p$ form. 
There were several implementation details that had to be taken into account when building the stack. In one hand, most protocols were developed to run independently and interfere as little as possible with the execution flow of the application. On the other hand, the consensus protocols offer the possibility to execute with blocking and non-blocking calls. To support this sort of functionality, SITAN needs the capability to keep context information about the protocol instances, manage efficiently multiple timeouts, and have several threads cooperating effectively.

\paragraph*{Protocol Context}
The stack needs to maintain several context data structures about the current and previous protocol executions. In addition, each protocol allows several instances to be run in parallel. To allow this, every instance is uniquely identified with a name composed by an application supplied label (e.g., a counter) concatenated with the protocol ID. Since SITAN tolerates Byzantine faults, it can happen that a process loses the ability to communicate for some time and later on reconnects, but in meanwhile the remaining process may have reached to a (consensus) decision. It is therefore essential that the correct processes save the decisions in the context data structure, so that when a process recovers it is able to learn the decided values. The library saves every results of consensus executions in a list for a while, until either it learns that no process will require the result or when there is the need to garbage collect the memory. 

\paragraph*{Timeouts}
Several protocols of the stack use a timeout to decide when to broadcast the internal state. Since there can be more than one protocol scheduling periodic tasks, this can affect the communications and result in the loss of messages, which would slow down the protocol executions. To avoid this, the protocols schedule the transmission of the messages in a more rational way. By default, communication protocols use a shorter timeout (of 1 ms), since these protocols require less computations and a shorter timeout does not flood the network with duplicate messages. As consensus protocols require more time to process the incoming messages (e.g., to check signatures), they only send messages when there is a state update or when a larger timeout is triggered (in this case 10ms). We observed that this approach reduces substantially the noise in the wireless medium and, as a result, less messages are lost.

\paragraph*{Threads} A \emph{communication thread} carries out message transmissions independently from the remaining threads in the protocol stack. It runs permanently in background, and when there is a timeout it transmits the messages that have accumulated. In particular, it forwards the messages produced locally and from the other processes, and periodically re-transmits messages that failed to reach their destination.

The \emph{group membership thread} is responsible for detecting new processes in the network and trigger events when a process is absent, in order to update the network graph and, if necessary, recalculate the sink. This thread runs independent from other threads and works in background by sending and receiving heartbeats. When a process fails to send a heartbeat within a predefined interval ($ 5 * timeout $) then the membership is updated. If the absent process was in the sink, then a new sink protocol instance starts. 

Consensus protocols run in a specific thread that is independent from the other threads. Each consensus protocol offers blocking and non-blocking operations. In the non-blocking mode, this lets the application proceed with the computation right after it invokes a consensus primitive. It also allows the parallel execution of several consensus instances. When choosing a non-blocking call, the application may indicate a callback function to be invoked when a decision is reached or the application can later consult the consensus execution state (and eventually get the result). In the blocking mode, it is possible to associate a timeout, to prevent the application from keeping on waiting beyond a certain interval.

%
%
%
%

\section{Evaluation of SITAN}

Two experimental environments were setup to assess the behavior of SITAN under different conditions. The first consisted of a realistic scenario with several mobile devices running a simple android application developed on top of the protocol stack. The second was based on the network simulator NS-3~\cite{NS3}, which allowed the execution of the protocols at a larger scale (up to 100 nodes). In addition, the devices were physically distributed accordingly to the following topologies: (i) a \textit{grid}, where processes are spread out at fixed positions in a grid. This scenario mimics for instance the execution of an application in a crowded classroom where the users are holding their mobile devices near to each other and stay still; (ii) a \textit{random}, where processes move randomly across a rectangular room. This scenario describes the execution of the application when the users are mobile, e.g., in a coffee break of a conference, with people walking randomly across the room and eventually stopping to talk to each other.

The experiments covered scenarios with all users behaving correctly and cases with  devices acting erroneously. In terms of the fault model, this corresponds to: \textit{fault-free}, when every process behaves as it is expected to; and \textit{Byzantine}, where the maximum number of Byzantine processes that can be tolerated exists in the system ($f = \lfloor \frac{n-1}{3} \rfloor$). In this situation, faulty processes  send wrong values (for the proposal, status, phase and identity) to attempt to disrupt the operation of the protocols. As consensus protocols can have distinct performance depending on the values that are proposed, we decided to include tests covering the two extreme situations: for best performance there are \textit{unanimous} proposals, where every correct process starts (binary/multivalued/vector) consensus with the same value; and for worst performance, there are \textit{divergent} proposals as every process picks a different value (except for the binary consensus where a process chooses 1 if its identifier is odd and 0 otherwise).

\subsection{Mobile devices environment}

The experimental evaluation of SITAN was performed in a platform composed by a group of 7 heterogeneous devices (emulating an ad hoc meeting): 
3 $\times$ Nexus 5 running Android 5.0 Lollipop;
2 $\times$ Asus Memo Pad 7 running Android 4.2 Jelly Bean;
2 $\times$ OnePlus One running Android 4.4 KitKat. The communication was carried out through WiFi Direct (without an access point), and so the distributed application exchanged messages in a pure $p2p$ form. 

The following questions were assessed with these experiments: (a) Is SITAN usable in real world? (b) Does the execution environment, the android platform, enable an acceptable performance of the protocol stack? (c) How differently do the protocols behave when running in optimal and adverse settings? 

\paragraph*{Optimal scenario} 
In the optimal execution scenario every process behaves as expected, and therefore no Byzantine faults occur during the experiment. The communication conditions are good (almost no messages are lost) and every process proposes the same value to the consensus protocols.

\begin{table}
\centering
\begin{tabular}{  c | c | c | c | c | c  }
  \hline                        
  \hline                        
  \textbf{N} & \textbf{Discovery}	& \textbf{Sink} & \textbf{Binary} & \textbf{Multivalued} & \textbf{Vector} \\
  \hline       
  \hline                        
  $ 4 $ & $ 57 \pm 7 $ & $ 103 \pm 22 $ & $ 53 \pm 8 $ & $ 155 \pm 23 $ & $ 186 \pm 19 $  \\
  \hline                        
  $ 5 $ & $ 91 \pm 11 $ & $ 106 \pm 27 $ & $ 79 \pm 12 $ & $ 157 \pm 39 $ & $ 216 \pm 16 $  \\
  \hline                        
  $ 6 $ & $ 146 \pm 19 $ & $ 131 \pm 22 $ & $ 102 \pm 7 $ & $ 191 \pm 15 $ & $ 281 \pm 24 $  \\
  \hline                        
  $ 7 $ & $ 198 \pm 21 $ & $ 198 \pm 35 $ & $ 177 \pm 14 $ & $ 282 \pm 30 $ & $ 337 \pm 20 $  \\
  \hline                        
  \hline                        
\end{tabular}
\caption{Average latencies and confidence interval under optimal conditions with $N$ devices (latency in milliseconds and confidence level of 95\%).}
\label{tab:exec_optimal}
\end{table}

In the experiments, SITAN was run with a set of 4 to 7 motionless devices in a grid (a small room with 6m $\times$ 6m). Table~\ref{tab:exec_optimal} shows the average latencies of 10 executions of the various protocols. We did not evaluate specifically the lowest level protocols, e.g., \textit{neighbors discovery}, because they run permanently in background to exchange messages, to detect missing processes and discover new ones.  The discovery and sink protocols execute in less than 200 ms. This means that after after a fifth of a second the protocol stack is initialized and the application can start to execute. The latency of every protocol increases with the number of processes, which is expected as having larger groups requires more messages to be exchanged and processed. Both multivalued consensus and vector consensus take a higher execution time when compared to binary consensus, but this is explained by need to perform several underlying consensus executions in order to reach a decision.

\paragraph*{Adverse scenario}
This scenario enables the assessment of SITAN under the conditions that approach the limits of the system model. Here, $f$ is set to the maximum tolerable number of faults by the protocols. Consequently, during the experiments $f = \lfloor \frac{n-1}{3} \rfloor$ processes behave unexpectedly by placing incorrect values in the messages. In addition, each process votes with divergent proposals. Finally, the devices are spread across three floors of a building, meaning that not every process is able to communicate directly with each other, and therefore messages need to be forward to reach the final destination.

\begin{table}
\centering
\begin{tabular}{  c | c | c | c | c | c | c  }
  \hline        
  \hline   
  \textbf{N} & \textbf{F} & \textbf{Discovery}	& \textbf{Sink} & \textbf{Binary} & \textbf{Multivalued} & \textbf{Vector} \\
  \hline      
  \hline   
  $ 4 $ & $ 1 $ & $ 74 \pm 8 $ & $ 122 \pm 22 $ & $ 66 \pm 8 $ & $ 203 \pm 17 $ & $ 213 \pm 27 $  \\
  \hline                        
  $ 5 $ & $ 1 $ & $ 107 \pm 12 $ & $ 137\pm 30 $ & $ 91 \pm 13 $ & $ 267 \pm 28 $ & $ 282 \pm 38 $  \\
  \hline                        
  $ 6 $ & $ 1 $ & $ 186 \pm 21 $ & $ 189 \pm 31 $ & $ 133 \pm 14 $ & $ 324 \pm 51 $ & $ 353 \pm 45 $  \\
  \hline                        
  $ 7 $ & $ 2 $ & $ 233 \pm 23 $ & $ 238 \pm 35 $ & $ 222 \pm 40 $ & $ 384 \pm 70 $ & $ 472 \pm $ 62 \\
  \hline  
  \hline   
\end{tabular}
\caption{Average latencies and confidence interval in the adverse scenario with $N$ devices where $F$ of them are Byzantine  (latency in milliseconds and confidence level of 95\%).}
\label{tab:exec_worst}
\end{table}

Table~\ref{tab:exec_worst} shows the average latencies of 10 executions of the protocols in the adverse conditions. The results are worse when compared with the optimal scenario. Depending on the protocol, latency is around 13\% to 41\% higher. This extra cost is acceptable given the advantage of tolerating Byzantine processes during the protocol execution and overcoming the network unreliability caused by the distribution of the devices. In any case, vector consensus ended up taking less than half a second.

Overall, experiments show that the protocols are able to finish their execution within a small time window, which is acceptable for various kinds of applications. Less favorable conditions have an impact on the observed performance, resulting in increased latencies, but we expect that this is a small penalty for the ability to communicate with partial connectivity and  tolerate various kinds of failures (in the nodes and the network).


\subsection{Simulation environment}
In order to evaluate the scalability of SITAN, we implemented the protocols and conducted several experiments using the NS-3 network simulator~\cite{NS3}. The simulation setup placed between 4 to 100  devices in a random position in a room with $ 30m \times 30m$. The devices were not static as they moved in random directions at a speed of 1m/s. Each experiment was repeated 10 times both in the fault-free and byzantine scenarios (with $f = \lfloor \frac{n-1}{3} \rfloor$). 
These experiments provide evidence that can help us answer the following research questions: (a) How does the performance scale to large groups of devices?; (b) How overloaded is the network medium with distinct group sizes?; (c) How many phases are necessary to complete the binary consensus?

\begin{figure}[!t]
  \minipage{0.32\textwidth}
  \subfigure[Binary consensus]{%
    \includegraphics[width=\linewidth]{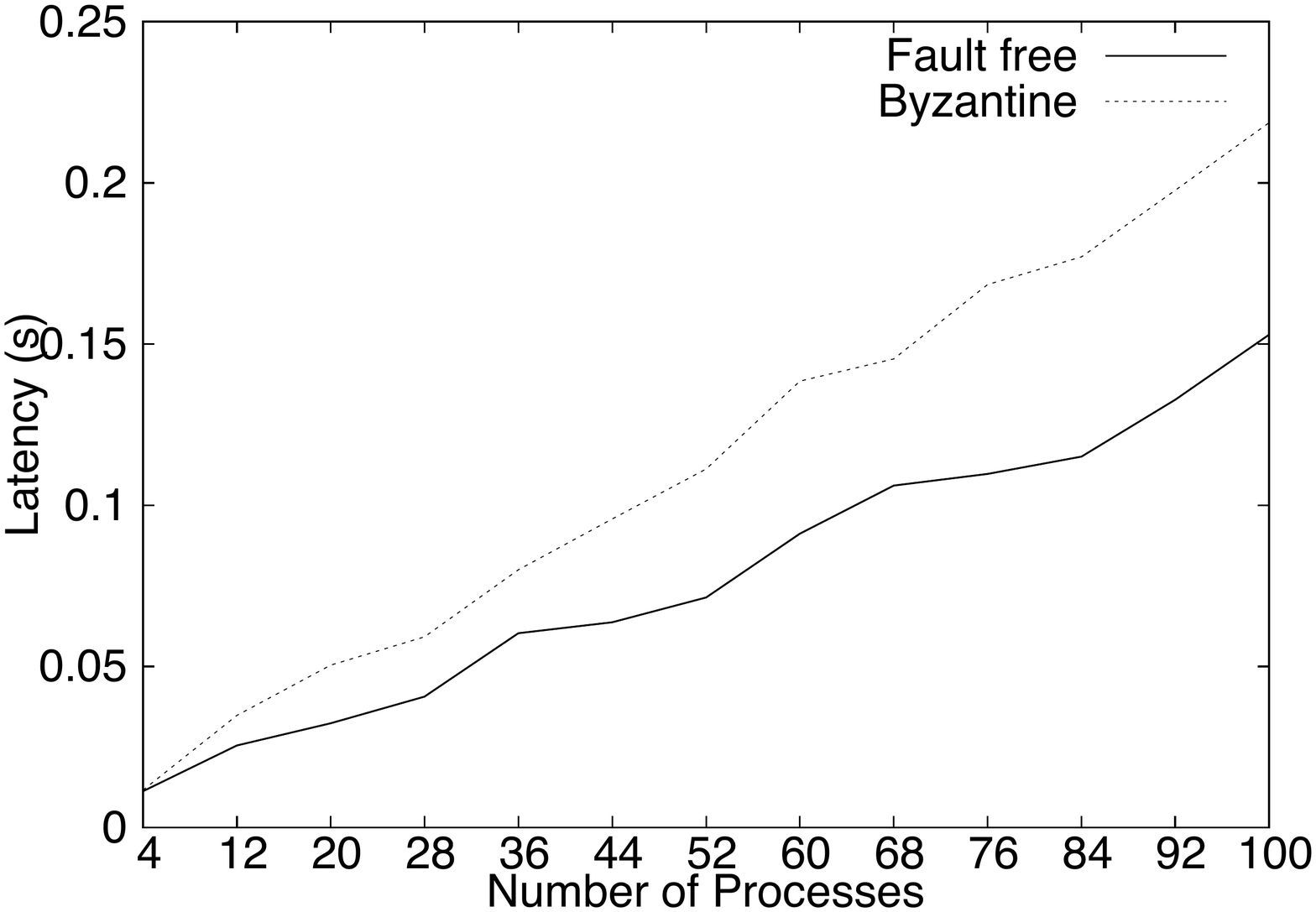}
    \label{ns3:latency_bc}
  }
  \endminipage\hfill
  \minipage{0.32\textwidth}
  \subfigure[Multivalued consensus]{%
    \includegraphics[width=\linewidth]{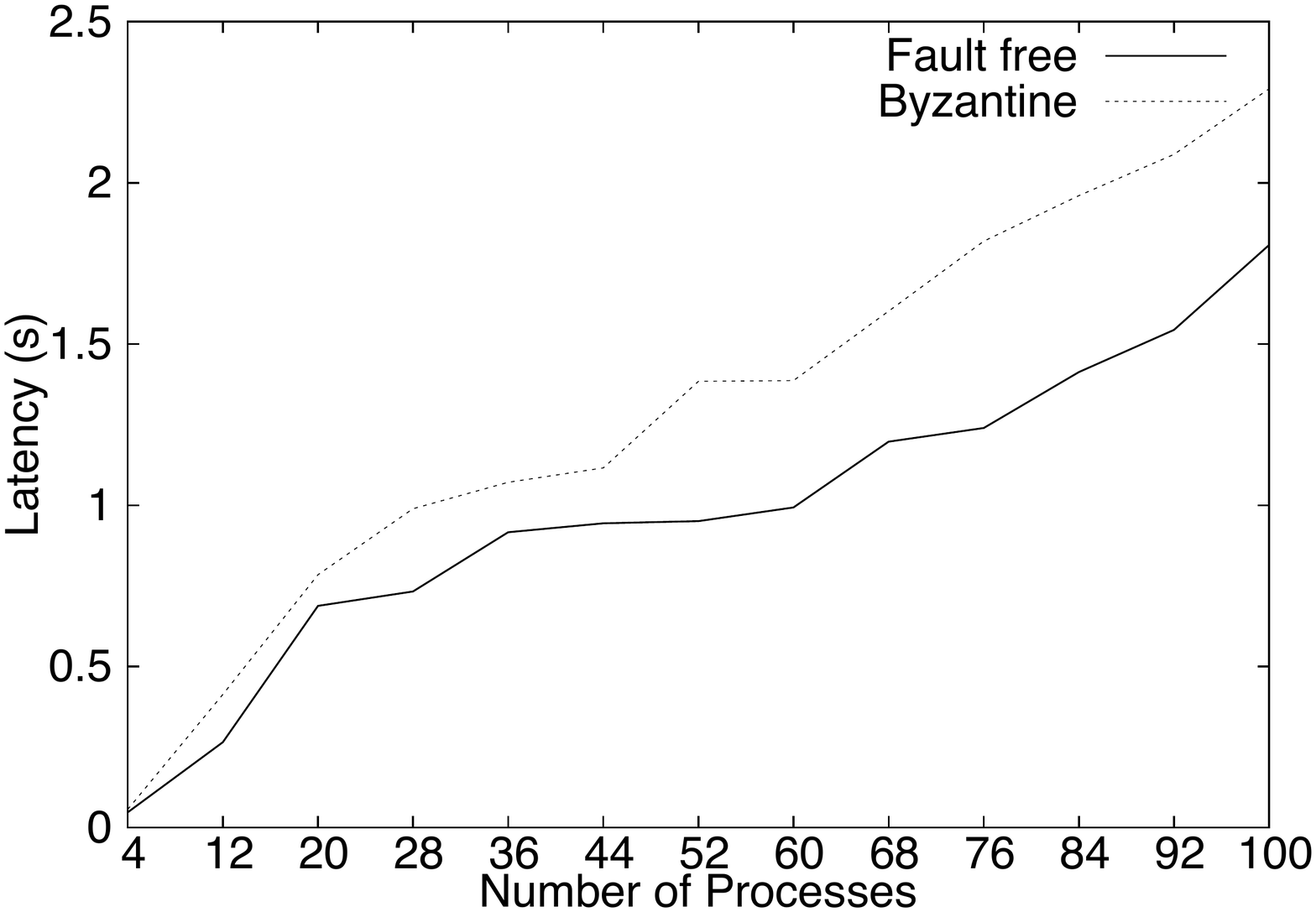}
    \label{ns3:latency_mc}
  }
  \endminipage\hfill
  \minipage{0.32\textwidth}%
  \subfigure[Vector consensus]{%
    \includegraphics[width=\linewidth]{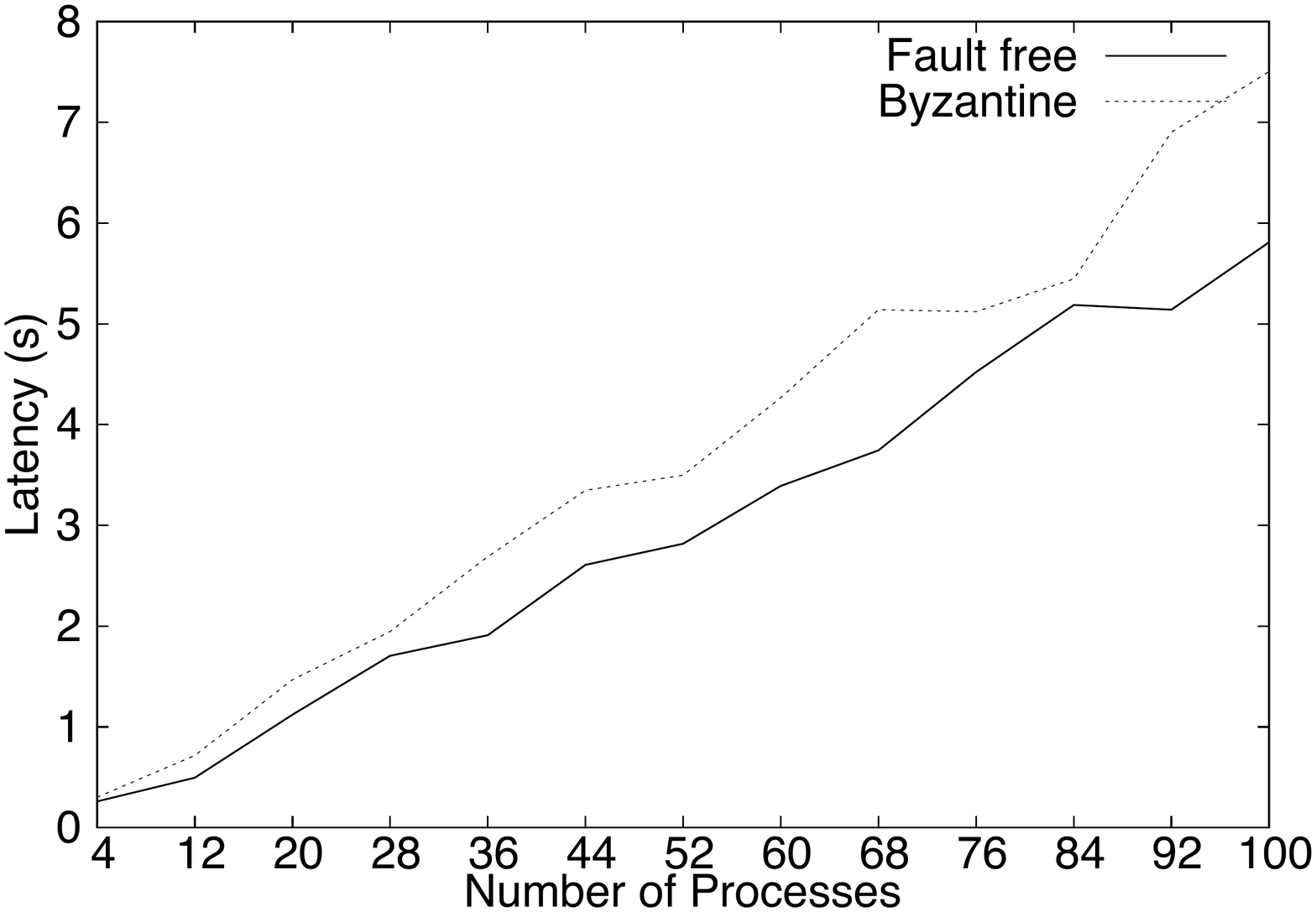}
    \label{ns3:latency_vc}
  }
  \endminipage
  \caption{Consensus latency in both fault-free and Byzantine scenarios.}
  \label{ns3:latency}
\end{figure}

Figure~\ref{ns3:latency} shows the latency of all three consensus protocols with unanimous proposals. In these experiments the sink network size varied from 4 to 16 processes. The latency increased as the group grew because more processes in the sink generate more messages to be processed. With 100 processes and failure-free conditions, latencies were in the order of 150ms, 1.8s and 5.8s for the binary, multivalued and vector consensus protocols respectively. Byzantine faults caused a limited increase of the overheads, since a decision could only be reached later on due to delays caused by messages with erroneous values. 

\begin{figure}
  \minipage{0.33\textwidth}
  \subfigure[Binary consensus]{%
    \includegraphics[width=\linewidth]{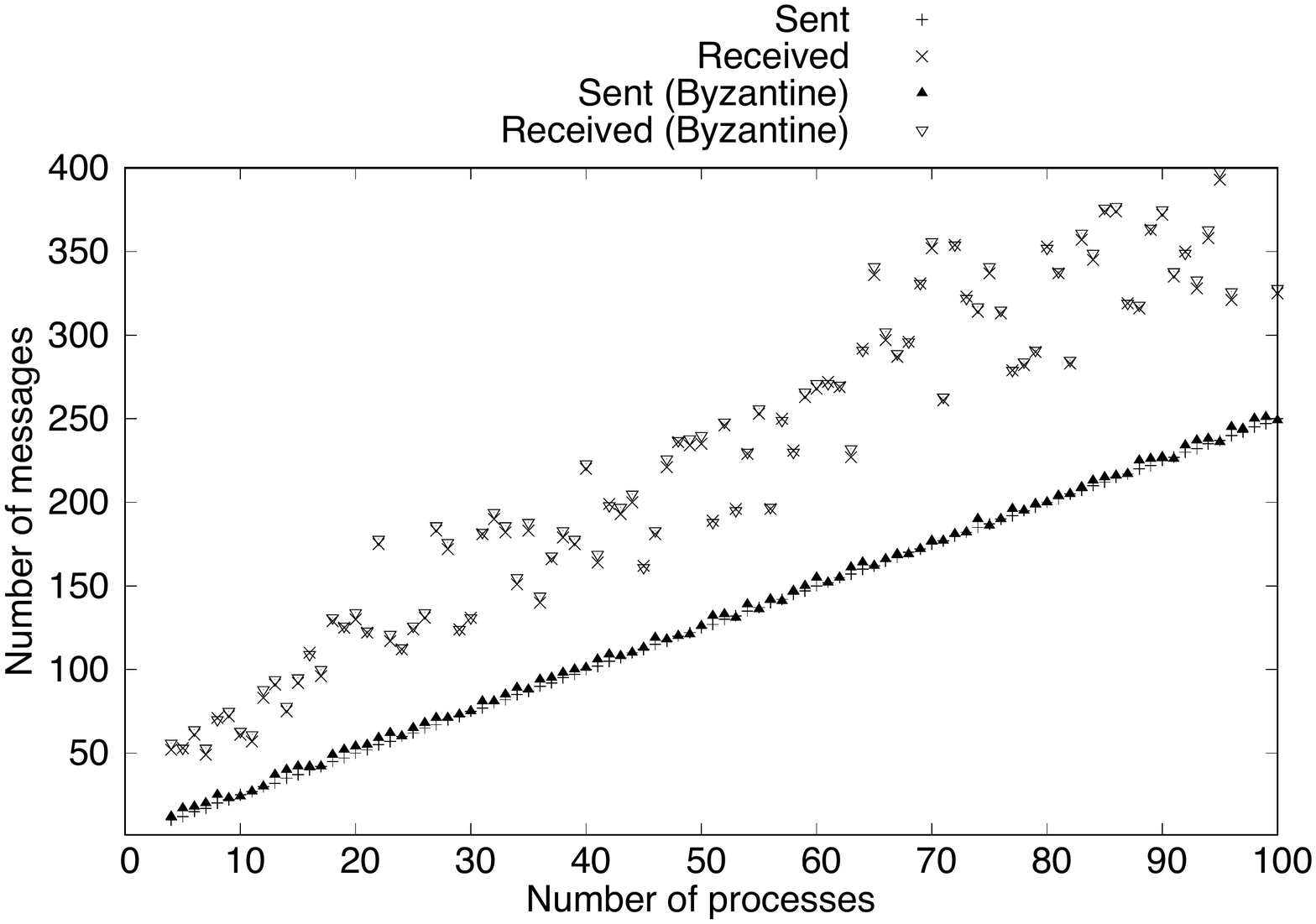}
    \label{ns3:msg_bc}
  }
  \endminipage\hfill
  \minipage{0.33\textwidth}
  \subfigure[Multivalued consensus]{%
    \includegraphics[width=\linewidth]{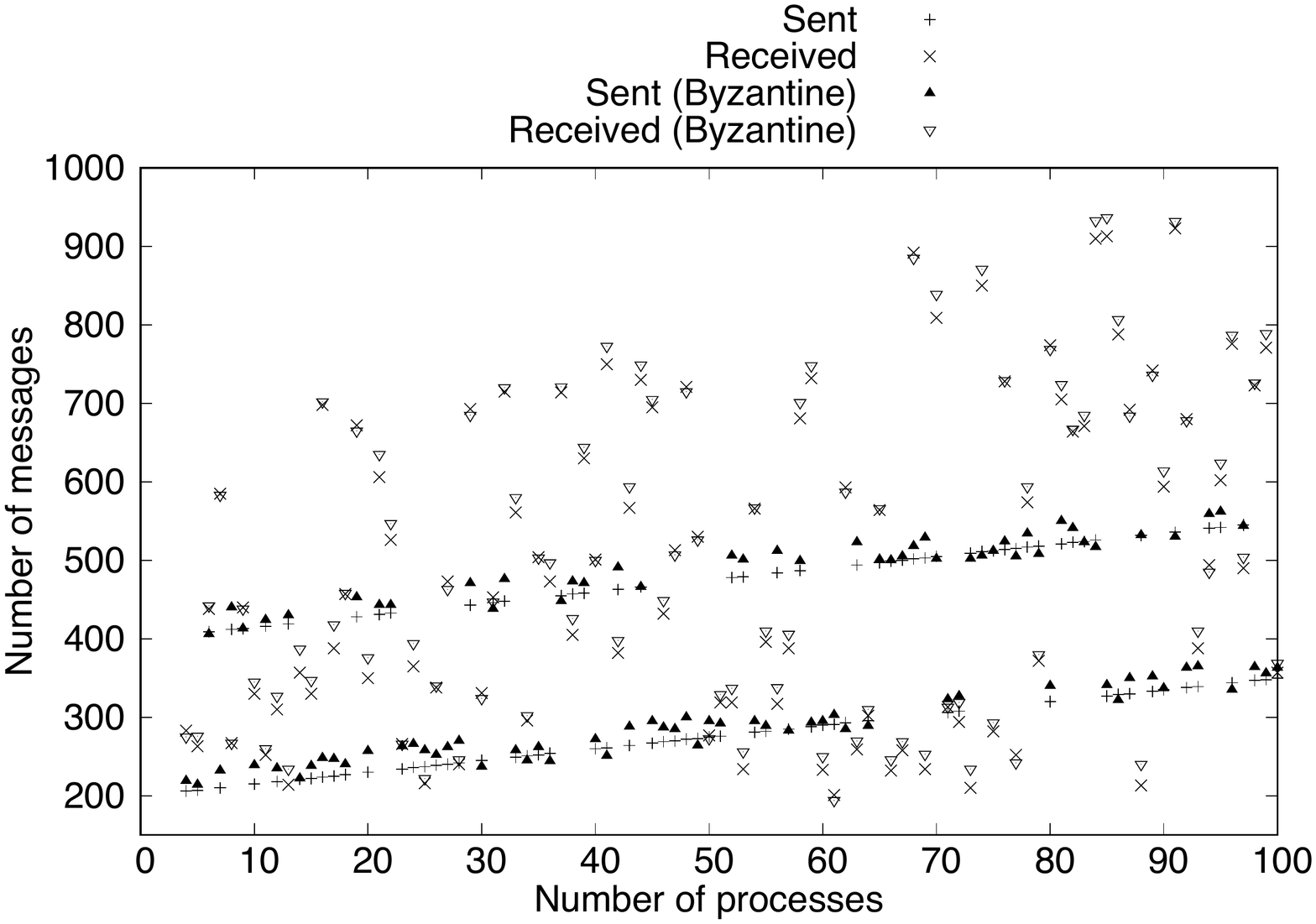}
    \label{ns3:msg_mc}
  }
  \endminipage\hfill
  \minipage{0.33\textwidth}%
  \subfigure[Vector consensus]{%
    \includegraphics[width=\linewidth]{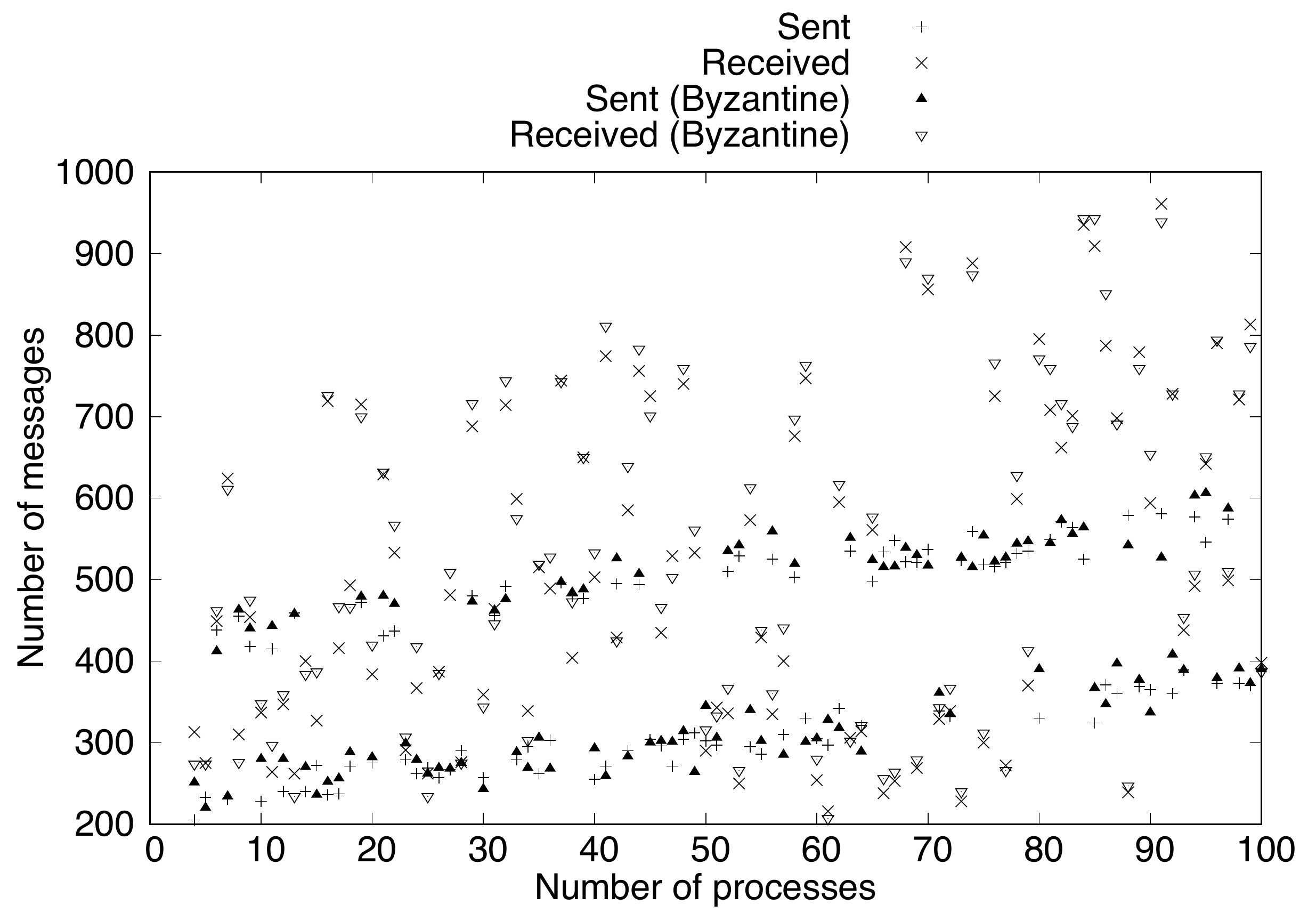}
    \label{ns3:msg_vc}
  }
  \endminipage
  \caption{Messages exchanged in a network with 4 -- 100 processes.}
  \label{ns3:msg}
\end{figure}

Figure~\ref{ns3:msg} shows the number of messages exchanged by the three consensus protocols. These values include all  messages transmitted by the underlying protocols, namely neighbors discovery, discovery and sink. In the binary consensus, the number of sends increases linearly with the number of processes (Figure~\ref{ns3:msg_bc}). There are three main effects that explain this result. First, more messages are required to discover and update the information about the members of larger groups, contributing to this aspect also the mobility of the devices. Second, as the sink grows, more processes transmit consensus related data. Lastly,  after a decision is reached, it is necessary to forward the result to higher numbers of devices that did not participate in the consensus. Multivalued and vector consensus also experience a rise on number of transmissions when the group grows (Figures~\ref{ns3:msg_mc} and~\ref{ns3:msg_vc}). However, this occurs at a much smaller pace, suggesting an acceptable level of scalability (e.g., in the vector consensus, the number of sends less than doubles for a 25$\times$ increase on the devices). 

\begin{figure}
\centering
\includegraphics[scale=0.45]{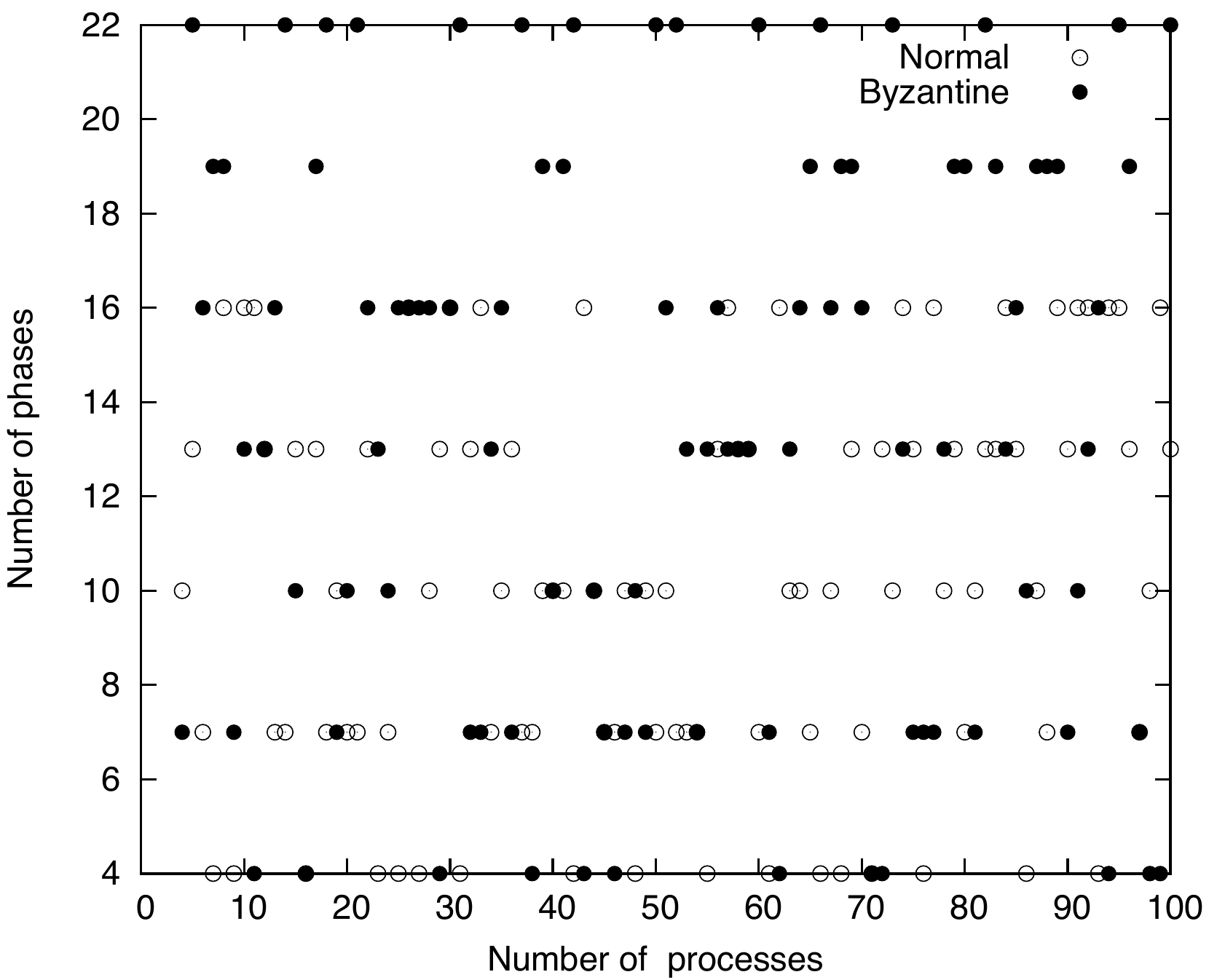}
\caption{Number of phases to achieve Binary Consensus}
\label{ns3:phases}
\end{figure}

Figure~\ref{ns3:phases} plots the number of phases it took to reach binary consensus with divergent proposals. These experiments run the protocol over the entire network, instead of using only the sink group. In our setting, the results show that the number of phases required to complete consensus does not increase with the number of processes. This means that up to 100 processes, the protocol does not seem to require more phases in order to converge. The results also indicate that when there are Byzantine processes sending erroneous messages, it can take at most 2 additional rounds for completion (of 3 phases each). This is an interesting outcome because it suggests that Byzantine processes end up being constrained on the amount of delay that they can inject.

\section{Conclusion}
This paper proposes a middleware with communication and agreement services for MANETs with unknown participants. SITAN is implemented by a number of protocols, allowing the configuration and management of the network and supporting the exchange of data and consensus among processes. These protocols were designed to address the main characteristics of MANETs, including unknown processes, unreliable communications, asynchrony and Byzantine failures. To the best of our knowledge, this is the first practical middleware that has all these characteristics.

Results show that it is possible to achieve binary consensus with 7 devices in the order of 200ms. Multivalued and vector consensus was completed in less than half a second.  Experiments with NS-3 show that SITAN is able to scale to bigger groups (up to 100 processes), which make these protocols adequate for reasonably sized MANETs. The Byzantine failures that were injected had a limited effect on the behavior of the protocols.

\section*{References}

\bibliography{mybibfile}


\end{document}